# Bias-field-free spin Hall nano-oscillators with an out-of-plane precession mode

Takanori Shirokura[1] and Pham Nam Hai[1,2,3*]

[1]*Department of Electrical and Electronic Engineering, Tokyo Institute of Technology,*

*2-12-1 Ookayama, Meguro, Tokyo 152-0033, Japan*

[2]*Center for Spintronics Research Network (CSRN), The University of Tokyo,*

*7-3-1 Hongo, Bunkyo, Tokyo 113-8656, Japan*

[3]*CREST, Japan Science and Technology Agency,*

*4-1-8 Honcho, Kawaguchi, Saitama 332-0012, Japan*

*Corresponding author: pham.n.ab@m.titech.ac.jp

**Spin Hall nano-oscillators (SHNOs) are promising candidates for new microwave oscillators with high durability due to a small driving current. However, conventional SHNOs with an in-plane precession (IPP) mode require a bias field for stable oscillations which is not favored in certain applications such as neuromorphic computing. Here, we propose and theoretically analyze a bias-field-free SHNO with an in-plane hard axis and an out-of-plane precession (OPP) mode by solving the Landau-Lifshitz-Gilbert (LLG) equation analytically and numerically. We derive formulas for driving currents and precession frequency, and show that they are in good agreement with numerical simulation results. We show that our proposed SHNOs can be driven by much smaller bias current than conventional spin torque nano-oscillators.**

**Introduction**

Spin-torque nano-oscillators (STNOs) capable of microwave frequency oscillations with a high Q factor are promising candidates for microwave generators [1] and neuromorphic computing [2]. STNOs utilize the spin-transfer torque (STT) to excite the precession of a magnetic layer, and this precession can be converted into microwaves by using the tunneling magnetoresistance (TMR) effect, the giant magnetoresistance (GMR) effect, or dipole emission [1,2,3,4,5,6,7,8]. Kiselev *et al.* first demonstrated STNOs by using Co/Cu/Co multilayers with in-plane easy axes, where the magnetization oscillates in the film plane, i.e. the in-plane precession (IPP) mode. High tunability of the oscillation frequency by controlling the current was also demonstrated [1]. However, an external magnetic field is required in those STNOs and can result in noises and additional costs, and a large STT driving current degrades the reliability of those STNOs. To solve these problems, the bias-field-free STNOs with an out-of-plane (OPP) mode were proposed using a perpendicular or tilting spin polarizer [4,5,6,7]. Furthermore, the bias-field-free STNO oscillating with a lower current was proposed by introducing the perpendicular magnetic anisotropy in the free magnetic layer. Nevertheless, the driving current was still as large as several tens to several hundred µA, which may cause long-term reliability problems [8].

For further reduction of the driving current, spin-Hall nano-oscillators (SHNOs) have been attracting much attention in recent years [9,10,11,12]. In the case of STNOs, the charge-to-spin current conversion efficiency is given by the spin-polarization *P* of the spin polarizing layer, which cannot exceed unity. Meanwhile, SHNOs utilize the spin Hall effect (SHE) for the spin current generation, and the charge-to-spin current conversion efficiency is given by $(L/t)\theta_{SH}$, where $L$ is the length of the SHNOs, $t$ is the thickness of the spin Hall layer, and $\theta_{SH}$ is the spin Hall angle of the spin Hall layer. Since $(L/t)\theta_{SH}$ can be larger than unity, the spin current can be generated more effectively in SHNOs than in STNOs. Furthermore, SHNOs possess higher durability because only pure spin current but no charge current is injected to the oscillating free magnetic layer. Several type of SHNOs with magnetic tunnel junctions (MTJs) [9], nano-wires [10], and nano-gaps [11,12] were proposed, and IPP mode microwave oscillations due to the spin-orbit torque (SOT) from the SHE were demonstrated. However, all those structures require an external magnetic field to sustain stable oscillations. In this paper, we propose and theoretically analyze a two-terminal bias-field-free SHNO with OPP mode by solving the LLG equation with macrospin approximation numerically and analytically. Our proposed SHNO has both the benefits of small driving current and bias-field free oscillation. We show the strategy for improving the performance of the bias-field-free SHNO. Our analysis can be applied as it is to three-terminal structures.

**Device structures**

Fig. 1 (a) and 1 (b) show two schematic device structures of our proposed two-terminal bias-field-free SHNO. In both structures, the MTJ is in contact with the spin Hall layer (spin source) by the free layer side. The spin Hall layer is composed of a material with a strong spin-orbit interaction such as a heavy metal or a topological insulator. The free layer has an in-plane hard axis parallel to the spin polarization vector of the spin current. The pinned layer also has an in-plane hard axis parallel to that of the free layer for maximizing the TMR effect. In SHE, the in-plane charge current in the spin Hall layer plays a crucial role because only the in-plane component of the charge current contributes to the pure spin current generation. In the SHNO with a parallel resistance shown in Fig. 1 (a), any spin source materials can be used since the parallel resistance provides the in-plane charge current path. Meanwhile, the perpendicular current flowing through the MJT converts the precession of the free layer to the microwave electrical signal. Here, we assume that the tunnel barrier of the MTJ is thick enough so that the perpendicular current is small and the STT effect from this current is negligible compared with the SOT effect. For the SHNO without the parallel resistance in Fig.1 (b), the in-plane current and the perpendicular current is the same. Therefore, a very thin spin Hall layer with a large spin Hall angle is required because the charge-to-spin current conversion efficiency is proportional to a spin Hall angle and inversely proportional to the thickness of the spin Hall layer. Topological insulators with two-dimensional surface states and giant spin Hall angle are suitable for the structure in Fig. 1(b). Note that the pinned layer is required only if one needs to extract the microwave electrical signal from the precession of the free layer. In applications such as microwave assisted magnetic recording (MAMR) that uses the microwave stray-field from the free layer, the tunnel barrier and the pinned layer can be omitted.

    The magnetization dynamics in the free layer is analyzed by solving the LLG equation with the antidamping-like SOT term [13,14],

$$\frac{d\bm{m}}{dt} = -\gamma \bm{m} \times \bm{H}_{\text{eff}} + \alpha \bm{m} \times \frac{d\bm{m}}{dt} - \gamma H_{\text{AD}} \bm{m} \times (\bm{m} \times \bm{p}_{\text{S}}),\tag{1}$$

where $\gamma$ is the gyromagnetic ratio, $\alpha$ is the Gilbert damping constant, $\bm{m}$ is the magnetization unit vector, $\bm{H}_{\text{eff}}$ is the effective magnetic field, $\bm{p}_{\text{S}}$ is the spin polarization unit vector. The strength of the spin-orbit torque is given by,

$$H_{\text{AD}} = \frac{\hbar \theta_{\text{SH}} I_{\text{C}}}{2eM_{\text{S}}t_{\text{FM}}t_{\text{SS}}W_{\text{SS}}},\tag{2}$$

where $\hbar$ is the Dirac's constant, $\theta_{\text{SH}}$ is the spin Hall angle, $I_{\text{C}}$ is the charge current in the spin source or the driving current, $e$ is the electronic charge, $M_{\text{S}}$ is the saturation magnetization of the free layer, $t_{\text{FM}}$ is the thickness of the free layer, $t_{\text{SS}}$ and $W_{\text{SS}}$ are the thickness and the width of the spin source, respectively. Here, we first ignore the field-like SOT term originating from the Rashba-Edelstein effect [15] because we want to focus on the antidamping-like SOT term originating from SHE. The general case with both the antidamping-like and field-like SOT term is discussed later. Fig. 1 (c) shows the schematic spin source / magnetic free layer and the coordination system for our simulation, where $W_{\text{FM}}$ and $L_{\text{FM}}$ are the width and the length of the free layer, respectively. Here, we assume $W_{\text{SS}} = W_{\text{FM}}$. When the current flows to the $x$ direction in the spin source, the spin current is injected to $-z$ direction with the spin polarization pointing toward the $-y$ direction. In this setup, we found that bias-free oscillation can be obtained under condition that the hard axis of the free layer is along the $y$ axis, namely, $N'_y > N'_x$, $N'_z$ should be satisfied, where $N'_x$, $N'_y$, and $N'_z$ are the effective anisotropic coefficients with respect to the $x, y, z$ direction, respectively. $N'_x$, $N'_y$, and $N'_z$ can be controlled by the shape of the free layer, the bulk crystalline anisotropy, the interfacial anisotropy, among others. To simplify the simulation, we assume that the effective anisotropic coefficients are controlled by only the shape of the free layer without the loss of generality. Table 1 shows the simulation parameters. We assume Tungsten (W) for the spin source material [16], and CoFeB for the free layer material [17]. The shape anisotropy is calculated by using the demagnetizing tensor of a rectangular shape [18]. We emphasize here that the size of the free layer with the rectangular shape assumed in Table 1 is for controlling the condition $N'_x$, $N'_z < N'_y$ poorly by the shape anisotropy, which is for the sake of simplicity and not suitable for realistic devices. In reality, the free layer should be a nanowire along the $x$ direction with the thickness of only a few nm to avoid the current shunting effect. In this case, we can obtain a large magnetic anisotropy constant $K_{uz}$ along the $z$ direction by utilizing the bulk crystalline magnetic anisotropy or the interfacial magnetic anisotropy. Similarly, we can obtain a large $K_{ux}$ by using uniaxial strain-induced magnetic anisotropy and field annealing-induced magnetic anisotropy, in addition to shape anisotropy along the $x$ direction. For example, $N'_x$ with the shape anisotropic coefficient $N_x$ and large $K_{ux}$ is given by $N'_x = N_x - K_{ux} / 2\pi M_S^2$.

**Numerical simulation and analytical analysis**

Fig. 2 (a) shows the time evolution of $\boldsymbol{m}$ with $L_{FM} = t_{FM} = 20$ nm and $W_{FM} = 15$ nm at the constant bias current of $I_C = 2$ μA. We observe the OPP stable precession of the magnetization around the $y$ axis without any bias field and at arbitrary small bias currents. However, such a precession requires the strict condition of $L_{FM} = t_{FM}$. Fig. 2 (b) and 2 (c) show the time evolution of $m$ when $L_{FM}$ is slightly changed to 21 nm at $I_C = 45$ µA and 46 µA, respectively. As shown in Fig. 2 (b) and 2 (c), the magnetization was relaxed to an equilibrium point ($I_C = 45$ µA) or immediately to the $-y$ direction ($I_C = 46$ µA) without sustainable oscillation. The same behaviors of the magnetization were observed at other $L_{FM} \neq t_{FM}$.

To understand these behaviors of the magnetization and find a way to obtain OPP stable oscillation in the general case of $L_{FM} \neq t_{FM}$, we analytically solve the time evolution of the energy of the magnetization [19,20]. In the following derivation, we consider the case of $N'_X < N'_Z$ ($t_{FM} < L_{FM}$) without the loss of generality. The magnetization energy has the minimum energy $E_{min}$ at the $x$-axis direction $m_o = (\pm 1, 0, 0)$, the maximum energy $E_{max}$ at the $y$-axis direction $m_{max} = (0, \pm 1, 0)$, and the saddle point energy $E_{sad}$ at the $z$-axis direction $m_{sad} = (0, 0, \pm 1)$. We found that OPP stable precession is obtained at $E_{max} > E > E_{sad}$, whose constant energy curves are shown by red curves in Fig. 3(a). This condition is consistent with that for STNOs oscillating in the OPP mode under a bias-field [20]. Next, we calculate the current needed for such OPP stable precession. The time evolution of the energy of the magnetization from the initial state $m_o = (\pm 1, 0, 0)$ to an arbitrary state is given by

$$\int_{t_0}^{t} dt \frac{dE}{dt} = W_{Re} + W_{SA} + W_{SF}. \tag{3}$$

Here, the work done by the damping torque $W_{Re}$, the antidamping-like torque $W_{SA}$, and the field-like torque $W_{\mathrm{SF}}$ are given by,

$$W_{\mathrm{Re}} = \gamma \alpha M_{\mathrm{S}} \int_{t_0}^{t} dt \left[ (\bm{m} \cdot \bm{H}_{\mathrm{eff}})^2 - H_{\mathrm{eff}}^2 \right], \tag{4}$$

$$W_{\mathrm{SA}} = \gamma M_{\mathrm{S}} H_{\mathrm{AD}} \int_{t_0}^{t} dt \left[ (\bm{m} \cdot \bm{p}_{\mathrm{S}})(\bm{m} \cdot \bm{H}_{\mathrm{eff}}) - \bm{p}_{\mathrm{S}} \cdot \bm{H}_{\mathrm{eff}} \right], \tag{5}$$

$$W_{\mathrm{SF}} = -\gamma \alpha M_{\mathrm{S}} H_{\mathrm{AD}} \int_{t_0}^{t} dt \bm{H}_{\mathrm{eff}} \cdot (\bm{m} \times \bm{p}_{\mathrm{S}}). \tag{6}$$

In a self-oscillation state, the average value of the time derivative of the energy for a precession period should be zero. Hence, the left-hand side of Eq. (3) becomes zero, and the current required to excite a self-oscillation on an arbitrary energy curve of $E_{\max} > E > E_{\mathrm{sad}}$ is given by,

$$I_C(E) = -\frac{2eM_{\mathrm{S}} t_{\mathrm{FM}} t_{\mathrm{SS}} W_{\mathrm{SS}}}{\hbar \theta_{\mathrm{SH}}} \frac{\alpha \oint dt \left[ (\bm{m} \cdot \bm{H}_{\mathrm{eff}})^2 - H_{\mathrm{eff}}^2 \right]}{\oint dt \left[ (\bm{m} \cdot \bm{p}_{\mathrm{S}})(\bm{m} \cdot \bm{H}_{\mathrm{eff}}) - \bm{p}_{\mathrm{S}} \cdot \bm{H}_{\mathrm{eff}} \right]}, \tag{7}$$

where the integral range is a precession period, and $W_{\mathrm{SF}}$ becomes zero in this integral range. Here, we assume that the magnetization precesses on the constant energy curve, although the actual trajectory of the magnetization has fluctuations around the constant energy curve. This approximation allows us to replace the time integral by the angle integral on the constant energy curve derived from much simpler damping-less LLG equation [21]. The integral of the numerator and the denominator in Eq. (7) are

$$\oint dt \left[ (\bm{m} \cdot \bm{H}_{\mathrm{eff}})^2 - H_{\mathrm{eff}}^2 \right] = -\frac{16\pi M_{\mathrm{S}} (N_y' - N_x') \sqrt{1-p^2}}{\gamma k} \left[ p^2 \mathrm{K}(\beta) + k^2 \{ \mathrm{E}(\beta) - \mathrm{K}(\beta) \} \right], \tag{8}$$

$$\oint dt \left[ (\bm{m} \cdot \bm{p}_{\mathrm{S}})(\bm{m} \cdot \bm{H}_{\mathrm{eff}}) - \bm{p}_{\mathrm{S}} \cdot \bm{H}_{\mathrm{eff}} \right] = \frac{2\pi p^2}{\gamma k}. \tag{9}$$

Therefore, Eq. (7) becomes

$$I_C(E) = \frac{16e\alpha M_S^2 t_{FM} t_{SS} W_{SS}(N_y' - N_x')}{\hbar \theta_{SH} p^2} \sqrt{1-p^2} \left[ p^2 K(\beta) + k^2 \{E(\beta) - K(\beta)\} \right], \quad (10)$$

where K($\beta$) and E($\beta$) are the first and second kinds of complete elliptic integral, respectively, $k$, $p$, and $\beta$ are defined as follows

$$k = \sqrt{\frac{N_y' - N_z'}{N_y' - N_x'}}, \quad (11)$$

$$p = \sqrt{\frac{2\pi M_S^2 N_y' - E}{2\pi M_S^2 N_y' - 2\pi M_S^2 N_x'}}, \quad (12)$$

$$\beta = \frac{p}{k} \sqrt{\frac{1-k^2}{1-p^2}}. \quad (13)$$

Then, the oscillation frequency is given by,

$$f(E) = \frac{\pi M_S \gamma k (N_y' - N_x') \sqrt{1-p^2}}{K(\beta)}. \quad (14)$$

In a self-oscillation state around the $y$ axis, the magnetization energy is larger than the saddle point energy $E_{sad}$ at the $z$-axis $\bm{m} = (0, 0, \pm 1)$, as shown in Fig. 3(a). Therefore, the minimum current required to sustain the precession around the $y$ axis is given by

$$I_{\min} \equiv \lim_{E \to E_{sad}} I(E) = \frac{16e\alpha M_S^2 t_{FM} t_{SS} W_{SS}(N_y' - N_x')}{\hbar \theta_{SH}} \sqrt{1-k^2}. \quad (15)$$

On the other hand, the current at which the magnetization is fully relaxed to –$y$ direction is given by

$$I_{\max} \equiv \lim_{E \to E_{\max}} I(E) = \frac{4\pi e\alpha M_S^2 t_{FM} t_{SS} W_{SS}(N_y' - N_x')}{\hbar \theta_{SH}} (1 + k^2). \quad (16)$$

There, the precession frequency takes a maximum value $f_{\max}$

$$f_{\max} \equiv \lim_{E \to E_{\max}} f(E) = 2M_S \gamma k \left( N'_y - N'_x \right). \tag{17}$$

To begin the precession around the *y* axis, the magnetization must first climb over the energy barrier $\Delta E$ between the initial point and the saddle point by the spin torque. The current required for magnetization to climb over $\Delta E$ is evaluated from Eq. (3) with the integral range from the initial point to the saddle point. In general, the trajectory of the magnetization between the initial point and the saddle point is complicated, and thus, it is difficult to solve the LLG equation analytically. Here, we approximate the trajectory between the initial state and the saddle point to the saddle energy curve, whose trajectory is shown by the blue curves in Fig. 3(a) [20]. In the case of $N'_x < N'_z$, the initial point $\boldsymbol{m_0} = (1, 0, 0)$ is replaced by the nearest point on the saddle energy curve $\boldsymbol{m_d} = (k, -\sqrt{1-k^2}, 0)$. Then, we obtain following equations,

$$\int_{t_0}^{t} dt \frac{dE}{dt} = E_{\text{sad}} - E_0, \tag{18}$$

$$\int_{t_0}^{t} dt \left[ (\boldsymbol{m} \cdot \boldsymbol{H}_{\text{eff}})^2 - \boldsymbol{H}_{\text{eff}}^2 \right] = -\frac{4\pi M_S \left( N'_y - N'_x \right)}{\gamma} k\sqrt{1-k^2}, \tag{19}$$

$$\int_{t_0}^{t} dt \left[ (\boldsymbol{m} \cdot \boldsymbol{p}_S)(\boldsymbol{m} \cdot \boldsymbol{H}_{\text{eff}}) - \boldsymbol{p}_S \cdot \boldsymbol{H}_{\text{eff}} \right] = \frac{\pi k}{2\gamma}, \tag{20}$$

$$\int_{t_0}^{t} dt \boldsymbol{H}_{\text{eff}} \cdot (\boldsymbol{m} \times \boldsymbol{p}_S) = -\frac{\sqrt{1-k^2}}{\gamma}. \tag{21}$$

By substituting Eq. (18) – (21) to Eq. (3) and solving for the current, we obtain

$$I_{\text{cri}} = \frac{8\pi e M_S^2 t_{\text{FM}} t_{\text{SS}} W_{\text{SS}} \left[ (N'_z - N'_x) + 2\alpha \left( N'_y - N'_x \right) k\sqrt{1-k^2} \right]}{\hbar \theta_{\text{SH}} \left( \pi k + 2\alpha \sqrt{1-k^2} \right)}. \tag{22}$$

$I_{\text{cri}}$ is composed of two terms; that needed for climbing over $\Delta E$ (the energy term: the first term inside

of the square brackets), and that for defeating the damping torque due to the demagnetizing field (the damping term: the second term inside of the square brackets). According to Eq. (22), the energy term has more influence on $I_{cri}$ than the damping term, since $\alpha$ is typically much smaller than unity. When $N'_Z < N'_X$, $N'_X$ and $N'_Z$ are exchanged in Eq. (8) – (22). Another expression of $I_{cri}$ was derived by using the spherical coordinate system as [22],

$$I_{cri}^{Sphe} = \frac{4\pi e M_S^2 t_{FM} t_{SS} W_{SS} (N'_Z - N'_X)}{\hbar \theta_{SH}} \tag{23}$$

In Eq. (23), only the energy term determines $I_{cri}$, while the damping term also affects $I_{cri}$ in our derived Eq. (22) (See the Supplementary Information for the influences caused by this difference.)

According to Eq. (15) and (22), $I_{cri}$ is always larger than $I_{min}$ at arbitrary $N'_X$, $N'_Z < N'_y$. Furthermore, $I_{cri}$ easily exceeds $I_{max}$ by a small difference between $L_{FM}$ and $t_{FM}$ because $I_{cri}$ is more sensitive to $\Delta E$ than $I_{max}$, as shown in Eq. (16) and (22). If we apply a constant current smaller than $I_{cri}$, the magnetization cannot climb over the saddle point. On the other hand, if we apply a constant current larger than $I_{cri}$, the magnetization will climb over the saddle point and then fully relax to the $y$-axis without any stable precession, consistent with the numerical simulation in Fig. 2(b) and 2(c). However, if we apply a very short pulse current $I_{pulse} > I_{cri}$ so that the magnetization climbs over the saddle point and then apply a much smaller constant current $I_{max} > I_{DC} > I_{min}$ after that, then stable OPP precession without fully relaxation can be realized even in the case of $L_{FM} \neq t_{FM}$. To check this assumption, we simulate the OPP precession by applying an initial short pulse current. Fig. 3 (b) and 3 (c) show the magnetic dynamics without and with the initial short pulse current at $L_{FM} = 21$ nm, respectively. The applied DC current in Fig. 3 (b) and (c) is 1.1 µA, which is larger than $I_{min}$. Without the short pulse current, the effective precession was not observed because the magnetization could not overcome $\Delta E$, as shown in Fig. 3 (b). On the other hand, the OPP mode precession around the $y$ axis was observed by applying the pulse current with the pulse amplitude of 46 µA and the pulse width of 1 ns. The rise time and the fall time of the pulse current were assumed to be 0 s. As shown in Fig. 3 (c), we can achieve the OPP stable precession even in $L_{FM} \neq t_{FM}$ if the pulse width is longer than the time required for the magnetization climbing over the saddle energy.

We checked the validity of Eq. (10) – (23) by solving the LLG equation numerically when varying $L_{FM}$ and $W_{FM}$. Fig. 4 (a) shows $L_{FM}$-dependence of $I_{cri}$ (blue), and Fig. 4 (b) shows that of $I_{max}$ (red), $I_{min}$ (green), and $f_{max}$ (orange), respectively. The solid lines show the analytical values given by Eq. (15), (16), (17), and (22), and the dots show the numerical simulation results. The orange dashed line in Fig. 4 (a) shows the analytical values given by Eq. (23). The simulation results of $I_{max}$, $I_{min}$, and $f_{max}$ in the region of $L_{FM} \geq 21$ nm were obtained with the pulse current excitation. $I_{cri}$ rapidly increased with increasing $L_{FM}$, and are much larger than $I_{max}$, as shown in Fig. 4 (a) and 4 (b). The analytical values are in good agreement with the numerical simulation results for $I_{min}$, $I_{max}$, $f_{max}$, and $I_{cri}$ at small $L_{FM}$. However, Eq. (22) overestimates $I_{cri}$ in the region of large $L_{FM}$, while Eq. (23) overestimates $I_{cri}$ for all $L_{FM}$. This difference between the simulation results and the analytical values of $I_{cri}$ given by Eq. (22) is attributed to the invalidity of the saddle energy curve approximation in the region of large $L_{FM}$ (see Supplementary Information). Fig. 4 (c) shows the current dependence of the precession frequency with several $L_{FM}$ ranging from 20 to 35 nm, where the solid lines are analytical values and dots are simulation results. The analytical values are also in good agreement with the simulation results. In the case of $L_{FM} \neq t_{FM}$, the precession frequency appears at $I_{min}$, and is roughly proportional to the current until saturation at $I_{max}$. This current dependence of the precession frequency is typical for the bias-field-free OPP mode [4,5,6,7]. Note that the precession frequency is strictly proportional to the current in the case of $L_{FM} = t_{FM}$, and the following useful relationship is obtained from Eq. (10) and Eq. (14)

$$f = \frac{\gamma \hbar \theta_{SH}}{4\pi e \alpha M_S t_{FM} t_{SS} W_{SS}} I_C. \tag{24}$$

**Analytical analysis in the general case with both the antidamping-like and field-like SOT term**

In this section, we discuss about the effect of the field-like SOT. The LLG equation with both the antidamping-like and the field-like SOT terms is given by

$$\frac{d\bm{m}}{dt} = -\gamma \bm{m} \times \bm{H}_{\text{eff}} + \alpha \bm{m} \times \frac{d\bm{m}}{dt} - \gamma H_{AD} \bm{m} \times (\bm{m} \times \bm{p}_S) - \gamma H_{FL} \bm{m} \times \bm{p}_S, \tag{25}$$

where, $H_{AD}$ and $H_{FL}$ are the strength of the antidamping-like and the filed-like SOT term, respectively. By rewriting the LLG equation in the form of the LL equation, we achieve the following expressions corresponding to Eq. (3) – (6) (see Supplementary Information).

$$\int_{t_0}^{t} dt \frac{dE}{dt} = W_{Re} + W_{SA} + W_{SF}, \tag{26}$$

$$W_{Re} = \gamma \alpha M_S \int_{t_0}^{t} dt \left[ (\bm{m} \cdot \bm{H}_{\text{eff}})^2 - \bm{H}_{\text{eff}}^2 \right], \tag{27}$$

$$W_{SA} = \gamma M_S (\alpha H_{FL} + H_{AD}) \int_{t_0}^{t} dt \left[ (\bm{m} \cdot \bm{p}_S)(\bm{m} \cdot \bm{H}_{\text{eff}}) - \bm{p}_S \cdot \bm{H}_{\text{eff}} \right], \tag{28}$$

$$W_{SF} = \gamma M_S (H_{FL} - \alpha H_{AD}) \int_{t_0}^{t} dt \bm{H}_{\text{eff}} \cdot (\bm{m} \times \bm{p}_S). \tag{29}$$

Because both $H_{AD}$ and $H_{FL}$ are proportional to a charge current $I_C$, we can solve Eq. (26) in the same manner as Eq. (3). By introducing the coefficients $\xi_{AD}$ and $\xi_{FL}$ defined by $H_{AD} = \xi_{AD} I_C$ and $H_{FL} = \xi_{FL} I_C$, respectively, we get the following formulas of $I_C(E)$, $I_{min}$, $I_{max}$, and $I_{cri}$ in the general case with both antidamping-like and fied-like SOT term.

$$I_C(E) = \frac{8\alpha M_S (N'_y - N'_x)}{(\alpha\xi_{FL} + \xi_{AD})p^2}\sqrt{1-p^2}\left[p^2 K(\beta) + k^2\{E(\beta) - K(\beta)\}\right], \tag{30}$$

$$I_{min} = \frac{8\alpha M_S (N'_y - N'_x)}{\alpha\xi_{FL} + \xi_{AD}}\sqrt{1-k^2}, \tag{31}$$

$$I_{max} = \frac{2\pi\alpha M_S (N'_y - N'_x)}{\alpha\xi_{FL} + \xi_{AD}}(1 + k^2), \tag{32}$$

$$I_{cri} = \frac{4\pi M_S\left[(N'_z - N'_x) + 2\alpha(N'_y - N'_x)k\sqrt{1-k^2}\right]}{\left(\alpha\pi k - 2\sqrt{1-k^2}\right)\xi_{FL} + \left(\pi k + 2\alpha\sqrt{1-k^2}\right)\xi_{AD}}. \tag{33}$$

Here, $\xi_{AD}$ is corresponding to $\dfrac{\hbar\theta_{SH}}{2eM_S t_{FM} t_{SS} W_{SS}}$. According to the Eq. (30), the effect of the field-like SOT term is small since $\alpha$ is typically much smaller than unity.

**Performance optimization**

The performance of our SHNO can be improved by optimizing the materials of the spin source and the free layer. Firstly, we demonstrate the improvement of the oscillation frequency. So far, we used only the shape anisotropy to control the effective anisotropic coefficients in the above simulations. In this case, the maximum oscillation frequency is limited by the saturation magnetization because the driving force of the precession is only the demagnetizing field. On the other hand, the oscillation frequency can be improved by using the uniaxial bulk crystalline anisotropy and the interfacial anisotropy. Fig. 5 (a) shows the relationship between $f_{max}$ and the uniaxial crystalline anisotropy $K_{ux}$ added to the $x$ direction with $W_{FM}$ ranging from 10 to 15 nm. Here, $f_{max}$ was calculated by using Eq. (17). From Fig. 5 (a), one can increase $f_{max}$ by introducing the uniaxial anisotropy. This can be understood from Eq. (14) and $N'_x = N_x - K_{ux}/2\pi M_S^2$. The dashed line in Fig. 5 (a) shows the line of $I_{max} = I_{min}$, namely, precession cannot be obtained on the right side of this line. However, the dashed line can be shifted to the right side by introducing anisotropy along the $z$ direction (for example, by perpendicular crystalline or interfacial magnetic anisotropy) because $I_{min}$ is decreased with decreasing the energy imbalances in the $x$–$z$ plane. Therefore, we can expand the precession region and obtain higher frequencies by controlling the magnetic anisotropy. Second, in MAMR application, there is a very large magnetic field of ~10 kOe in the gap between the main pole and the trailing shield of the write-head. If this field is applied to the $+y$ direction, the frequency can be increased because this external magnetic field disturbs magnetizing to the $-y$ direction. In this case, $f_{max}$ is improved by

$$f'_{max} = f_{max} + \frac{\gamma}{2\pi} H^y_{ext}, \qquad (25)$$

or 2.8 GHz per 1 kOe, where $H^y_{ext}$ is the external magnetic field applied to the $+y$ direction. Thus, high frequency of 25 GHz or higher is very obtainable for MAMR applications with the write gap field of 10 kOe using our OPP mode SHNO. Furthermore, $I_{cri}$ is also reduced by applying small $H^y_{ext}$ because $\Delta E$ is reduced with decreasing the in-plane component of $m_0$ and $m_{sad}$, namely, the energy term in Eq. (22) is reduced. However, large $H^y_{ext}$ increases $I_{cri}$ because the damping term in Eq. (22)

is increased, and thus, $I_{cri}$ takes a minimum value with respect to $H^y_{ext}$ due to the trade-off between the energy term and the damping term in Eq. (22).

Finally, we demonstrate the reduction of the driving current. In our simulations, we assumed W as the material of the spin source. Although W possesses the largest $\theta_{SH} \sim 0.4$ among heavy metals, much larger $\theta_{SH}$ can be obtained by utilizing topological insulators. Fig. 5 (b) shows the comparison of the driving current for 3 GHz oscillation using heavy metals (W, Pt, Ta) and topological insulators ($Bi_xSe_{1-x}$, BiSb(012)) for the spin source [16,23,24,25,26,27]. The structure of the SHNO is assumed $W_{SS} : t_{SS}$ = 15 nm : 2 nm for the spin source, and $L_{FM} : W_{FM} : t_{FM}$ = 20 nm : 15 nm : 20 nm for the free layer. For comparison, we also simulated the bias-field-free OPP mode STNO with a perpendicular spin polarizing layer. The structure of the STNO is assumed $L_{FM} : W_{FM} : t_{FM}$ = 50 nm : 50 nm : 2 nm as the free layer whose volume is almost same as that of the SHNO. The material for the spin polarizing layer of the STNO is assumed to be CoFeB [17]. From Fig. 5 (b), the driving currents of SHNOs with heavy metals are already smaller than that of the STNO. Furthermore, the driving currents of SHNOs with topological insulators are very small, at the order of 10 nA. These ultra-small driving currents improve long-term durability of the SHNO. In the case of topological insulators, their high resistivity is a problem [25,28]. Therefore, using topological insulators with high conductivity and large spin Hall angle, such as BiSb, is necessary for SHNOs [26,27].

**Conclusion**

In conclusion, we propose SHNOs that can oscillate in OPP mode without a bias field, by designing the hard axis of the magnetic free layer parallel to the spin polarized direction of the pure spin current. The oscillation can be excited by applying the initial short pulse current $I_{cri}$ at arbitrary $N'_x$, $N'_z < N'_y$, followed by a small DC current $I_{max} > I_{DC} > I_{min}$. We derived analytical equations of $I_{cri}$, $I_{max}$, $I_{min}$, $f_{max}$ for the SHNOs. Furthermore, we show that the oscillation frequency can be increased by controlling the magnetic anisotropy, or by applying a magnetic field to the $+y$ direction. We show that the driving current of our SHNOs can be significantly reduced by using topological insulators as the spin Hall material. Our SHNOs are promising for various microwave applications, such as neuromorphic computing or MAMR with long-term durability.

**Supplementary Material:** See the supplementary material for section I: Derivation of Eq. (10), (14), and (22), section II: Effect of the field-like SOT, and section III: Critical current.

**Acknowledgment**: This work was supported by JST-CREST (JPMJCR18T5).

**Table 1.** Simulation parameters.

| Parameters | Values | Parameters | Values |
| --- | --- | --- | --- |
| $W_{SS}$ [nm] | 10 ~ 19 | $M_S$ [emu/cc] | 1200 |
| $t_{SS}$ [nm] | 2 | $L_{FM}$ [nm] | 20 ~ 35 |
| $\theta_{SH}$ | 0.4 | $W_{FM}$ [nm] | 10 ~ 19 |
| $\alpha$ | 0.005 | $t_{FM}$ [nm] | 20 |

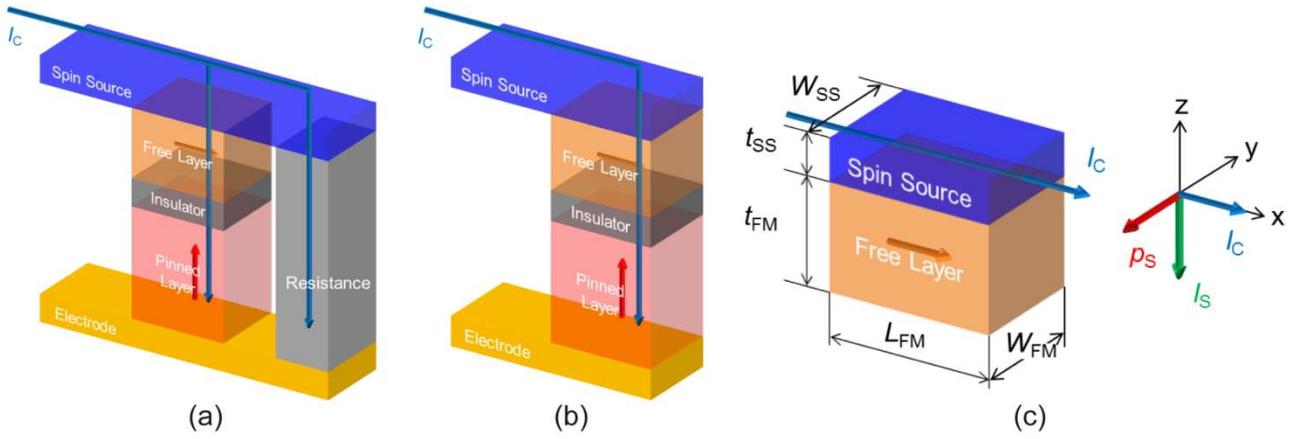

**Fig. 1.** Schematic structure of two-terminal bias-field-free SHNOs **(a)** with a parallel resistance, and **(b)** without a parallel resistance. **(c)** Coordination system and simulation parameters. The orange and red arrows show the magnetization of the free and pinned layer, respectively. The blue arrows show the current flow paths.

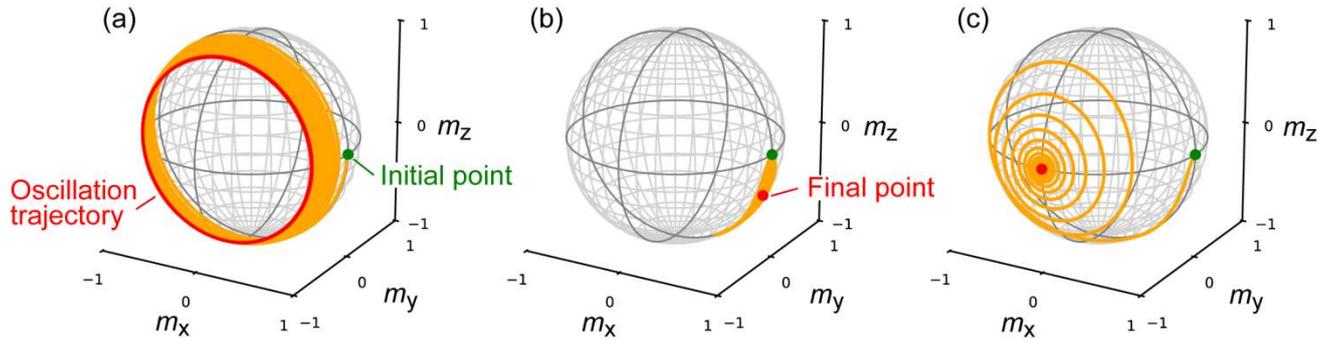

**Fig. 2.** Time evolution of the magnetization unit vector *m* when **(a)** $L_{FM} = W_{FM} = 20$ nm and $t_{FM} = 15$ nm at $I_C = 2$ µA, **(b)** $L_{FM}$ is changed to 21 nm at $I_C = 45$ µA, and **(c)** $L_{FM} = 21$ nm at $I_C = 46$ µA. The green points, the orange solid lines, the red solid line, and the red points show the initial points, the trajectories of the magnetization, the oscillation trajectory in a self-oscillation state, and the final points, respectively.

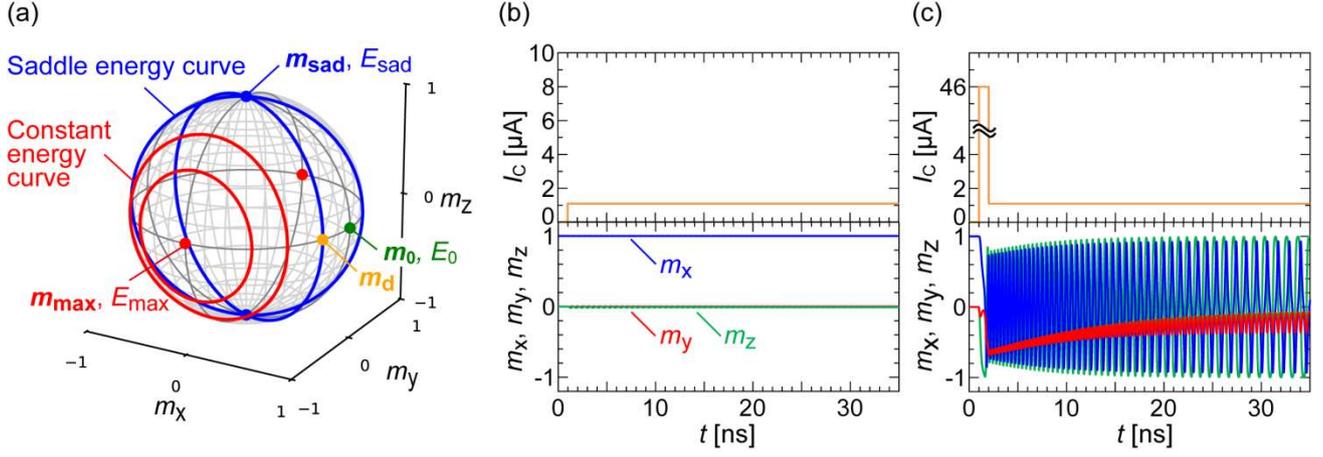

**Fig. 3. (a)** Schematic illustration of the constant energy curves and important points in the magnetization unit vector space. Green $m_0$ and $E_0$ indicate the initial point and the minimum energy. Blue $m_{sad}$ and $E_{sad}$ indicate the saddle point and the saddle energy curves. Orange $m_d$ indicates the nearest point to $m_0$ in the saddle energy curves. Red $m_{max}$ and $E_{max}$ indicate the maximum point and maximum energy. Red curves indicate constant energy curves for $E_{max} > E > E_{sad}$. Time evolution of the current and the magnetization with **(b)** only a DC current of 1.1 μA and no initial pulse current, and **(c)** with a 46 μA pulse current excitation followed by a DC current of 1.1 μA.

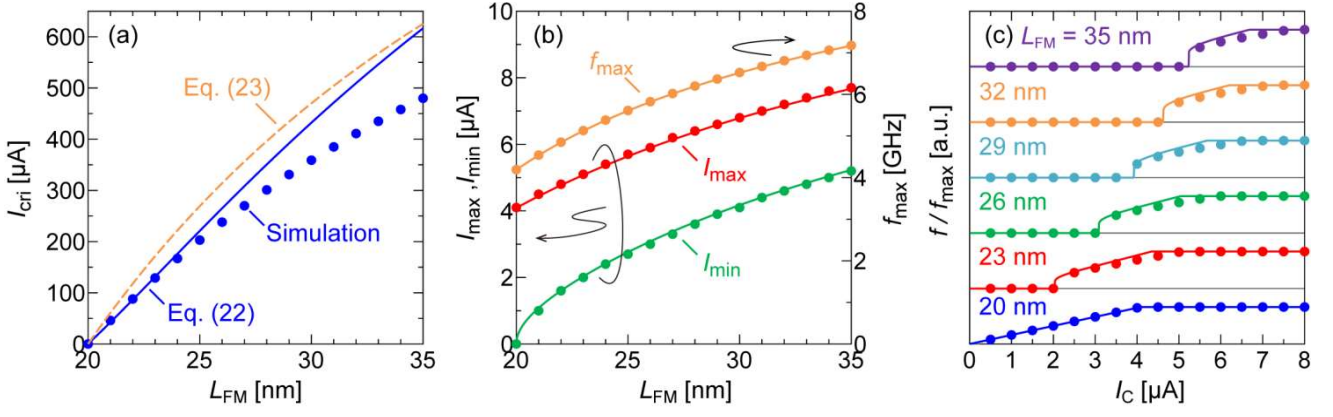

**Fig. 4.** $L_{FM}$-dependence of **(a)** the critical current $I_{cri}$, **(b)** the maximum current $I_{max}$ (red), the minimum current $I_{min}$ (green), and the maximum precession frequency $f_{max}$ (orange), respectively. **(c)** $I_C$-dependence of the normalized frequency $f / f_{max}$ with $L_{FM}$ ranging from 20 to 35 nm. The solid and dashed lines show the analytical results, and the dots show the numerical simulation results.

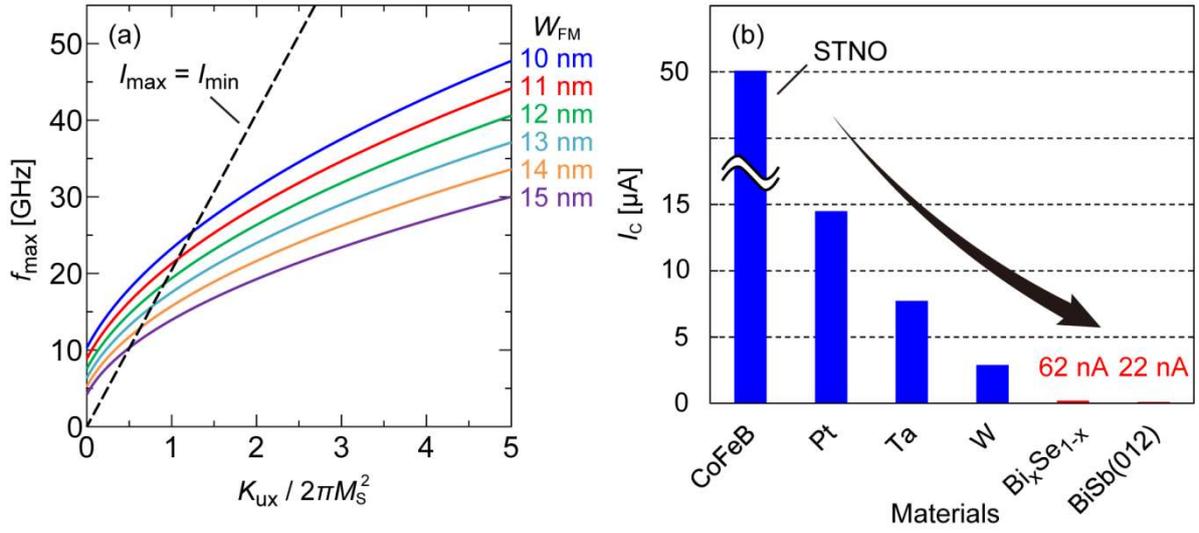

**Fig. 5. (a)** Maximum precession frequency $f_{max}$ as a function of the uniaxial crystalline anisotropy $K_{ux}$ added to the $x$ direction with the free layer $W_{FM}$ ranging from 10 to 15 nm. The dashed line shows the condition $I_{max} = I_{min}$. **(b)** Spin source material dependence of the driving current $I_C$ of our SHNOs. $I_C$ of a STNO with CoFeB as the spin polarizing layer is also shown for comparison.

**References**

[1] S. I. Kiselev, J. C. Sankey, I. N. Krivorotov, N. C. Emley, R. J. Schoelkopf, R. A. Buhrman, and D. C. Ralph, Nature **425**, 308 (2003).

[2] J. Torrejon, M. Riou, F. A. Araujo, S. Tsunegi, G. Khalsa, D. Querlioz, P. Bortolotti, V. Cros, K. Yakushiji, A. Fukushima, H. Kubota, S. Yuasa, M. D. Stiles, and J. Grollier, Nature **547**, 428 (2017).

[3] O. Prokopenko, E. Bankowski, T. Meitzler, V. Tiberkevich, and A. Slavin, Magn. Lett., IEEE **2**, 3000104 (2011).

[4] X. Zhu and J. G. Zhu, IEEE Trans. Magn. **42**, 2670 (2006).

[5] D. Houssameddine, U. Ebels, B. Delaet, B. Rodmacq, I. Firastrau, F. Ponthenier, M. Brunet, C. Thirion, J.-P. Michel, L. Perjbeanu-Buda, M.-C. Cyrille, O. Redon, and B. Dieny, Nature Mater. **6**, 447 (2007).

[6] Y. Zhou, H. Zhang, Y. Liu, and J. Åkerman, J. Appl. Phys. **112**, 063903 (2012).

[7] G. Lv, H. Zhang, X. Cao, Y. Liu, Z. Hou, Y. Qin, G. Li, and L. Wang, AIP Adv. **5**, 077171 (2015).

[8] Z. Zeng, G. Finocchio, B. Zhang, P. K. Amiri, J. A. Katine, I. N. Krivorotov, Y. Huai, J. Langer, B. Azzerboni, K. L. Wang, and H. Jiang, Sci. Rep. **3**, 1426 (2013).

[9] L. Liu, C.-F. Pai, D. C. Ralph, and R. A. Buhrman, Phys. Rev. Lett. **109**, 186602 (2012).

[10] Z. Duan, A. Smith, L. Yang, B. Youngblood, J. Lindner, V. E. Demidov, S. O. Demokritov, and I. N. Krivorotov, Nat. Commun. **5**, 5616 (2014).

[11] V. E. Demidov, S. Urazhdin, H. Ulrichs, V. Tiberkevich, A. Slavin, D. Baither, G. Schmitz, and S. O. Demokritov, Nat. Mater. **11**, 1028 (2012).

[12] V. E. Demidov, S. Urazhdin, A. Zholud, A. V. Sadovnikov, and S. O. Demokritov, Appl. Phys. Lett. **105**, 172410 (2014).

[13] J. C. Slonczewski, J. Magn. Magn. Mater. **159**, L1 (1996).

[14] L. Berger, Phys. Rev. B **54**, 9353 (1996).

[15] V. M. Edelstein, Solid State Commun. **73**, 233 (1990).

[16] C.-F. Pai, L. Liu, Y. Li, H. W. Tseng, D. C. Ralph, and R. A. Buhrman, Appl. Phys. Lett. **101**, 122404 (2012).

[17] S. Ikeda, K. Miura, H. Yamamoto, K. Mizunuma, H. D. Gan, M. Endo, S. Kanai, J. Hayakawa, F. Matsukura, and H. Ohno, Nature Mater. **9**, 721 (2010).

[18] A. Aharoni, J. Appl. Phys. **83**, 3432 (1998).

[19] T. Taniguchi, Phys. Rev. B **91**, 104406 (2015).

[20] T. Taniguchi and H. Kubota, Phys. Rev. B **93**, 174401 (2016).

[21] G. Bertotti, I. Mayergoyz, and C. Serpico, *Nonlinear Magnetization Dynamics in Nanosystems* (Elsevier, Oxford, 2009), Chap. 4.

[22] R. Matsumoto and H. Imamura, AIP Adv. **6**, 125033 (2016).

[23] L. Liu, T. Moriyama, D. C. Ralph, and R. A. Buhrman, Phys. Rev. Lett. **106**, 036601 (2011).

[24] L. Liu, C.-F. Pai, Y. Li, H. W. Tseng, D. C. Ralph, and R. A. Buhrman, Science **336**, 555 (2012).

[25] Dc. Mahendra, R. Grassi, J.-Y. Chen, M. Jamali, D. Reifsnyder Hickey, D. Zhang, Z. Zhao, H. Li, P. Quarterman, Y. Lv, M. Li, A. Manchon, K. A. Mkhoyan, T. Low, and J.-P. Wang, Nature Mater. **17**, 800 (2018).

[26] Y. Ueda, N. H. D. Khang, K. Yao, and P. N. Hai, Appl. Phys. Lett. **110**, 062401 (2017).

[27] N. H. D. Khang, Y. Ueda, and P. N. Hai, Nature Mater. **17**, 808 (2018).

[28] A. R. Mellnik, J. S. Lee, A. Richardella, J. L. Grab, P. J. Mintun, M. H. Fischer, A. Vaezi, A. Manchon, E.-A. Kim, N. Samarth, and D. C. Ralph, Nature **511**, 449 (2014).

# Supplementary Information

**Bias-field-free spin Hall nano-oscillators with an out-of-plane precession mode**

Takanori Shirokura[1] and Pham Nam Hai[1,2,3]*

[1]Department of Electrical and Electronic Engineering, Tokyo Institute of Technology,

2-12-1 Ookayama, Meguro, Tokyo 152-0033, Japan

[2]Center for Spintronics Research Network (CSRN), The University of Tokyo,

7-3-1 Hongo, Bunkyo, Tokyo 113-8656, Japan

[3]CREST, Japan Science and Technology Agency,

4-1-8 Honcho, Kawaguchi, Saitama 332-0012, Japan

*Corresponding author: pham.n.ab@m.titech.ac.jp

# I. Derivation of Eq. (10), (14), and (22)

In this section, we derive Eq. (10), (14), and (22) under the assumption of $N'_X \leq N'_Z$. The equations under the condition of $N'_Z \leq N'_X$ can be obtained by in the same manner. When a magnetization $\boldsymbol{m}$ precesses on a constant energy curve of $E$, the following Eq. (S1) and (S2) are satisfied

$$m_x^2 + m_y^2 + m_z^2 = 1, \tag{S1}$$

$$E = 2\pi M_S^2 N'_X m_x^2 + 2\pi M_S^2 N'_y m_y^2 + 2\pi M_S^2 N'_z m_z^2. \tag{S2}$$

Here, Eq. (S1) is the definition of the magnetization unit vector, and Eq. (S2) is the conservation of the energy under the zero-bias-field. By substituting Eq. (S2) to Eq. (S1), we obtain the following equations

$$m_x^2 + k^2 m_z^2 = p^2, \tag{S3}$$

$$m_x = p\cos u, \tag{S4}$$

$$m_z = \frac{p}{k}\sin u, \tag{S5}$$

where $u$ as a function of time is the angle with respect to the $x$ axis in the $x$-$z$ plane, and $k$, $p$ are defined in Eq. (11) and (12), respectively. On the other hand, the dynamics of the magnetization on a constant energy curve is given by the damping-less LLG equation, and thus, the infinitesimal time $dt$ is given by,

$$dt = \frac{du}{4\pi M_S \gamma k (N'_y - N'_x) m_y} = \frac{-1}{4\pi M_S \gamma k (N'_y - N'_x)\sqrt{1-p^2}} \frac{du}{\sqrt{1-\beta^2 \sin^2 u}}. \tag{S6}$$

Eq. (14) can be obtained by integrating Eq. (S6) with the integration range of the one precession period,

$$f(E) = \frac{1}{\oint dt} = \frac{-\pi M_S \gamma k \left(N'_y - N'_x\right)\sqrt{1-p^2}}{\int_0^{-\frac{\pi}{2}} du \left(1 - \beta^2 \sin^2 u\right)^{-\frac{1}{2}}}. \tag{S7}$$

Eq. (10) can be obtained by replacing the integration variable of $t$ by $u$ in Eq. (7) by using Eq. (S6), as follows

$$\oint dt\left[(\mathbf{m}\cdot\mathbf{H}_{\text{eff}})^2 - \mathbf{H}_{\text{eff}}^2\right] = -(4\pi M_S)^2 \left(N'_y - N'_x\right)^2 p^2 \left(1-p^2\right)\oint dt\left(1 - \frac{k^2}{p^2}\beta^2 \sin^2 u\right)$$

$$= \frac{16\pi M_S \left(N'_y - N'_x\right)}{\gamma k}\sqrt{1-p^2} \int_0^{-\frac{\pi}{2}} du \left(\frac{p^2 - k^2}{\sqrt{1-\beta^2 \sin^2 u}} + k^2\sqrt{1-\beta^2 \sin^2 u}\right), \tag{S8}$$

$$\oint dt\left[(\mathbf{m}\cdot\mathbf{p}_S)(\mathbf{m}\cdot\mathbf{H}_{\text{eff}}) - \mathbf{p}_S\cdot\mathbf{H}_{\text{eff}}\right] = -4\pi M_S \left(N'_y - N'_x\right)p^2 \oint dt\, m_y = -\frac{4p^2}{\gamma k}\int_0^{-\frac{\pi}{2}} du, \tag{S9}$$

$$\oint dt\, \mathbf{H}_{\text{eff}}\cdot(\mathbf{m}\times\mathbf{p}_S) = 4\pi M_S \left(N'_z - N'_x\right)\frac{p^2}{k}\oint dt\, \cos u \sin u = 0. \tag{S10}$$

Eq. (22) also can be obtained by the same way of integration variable replacement. Because $p = k$ and $\beta = 1$ are held on the saddle energy curve, the integration of Eq. (19), (20) and (21) becomes very simple as shown below

$$\int_{t_0}^{t} dt\left[(\mathbf{m}\cdot\mathbf{H}_{\text{eff}})^2 - \mathbf{H}_{\text{eff}}^2\right] = -(4\pi M_S)^2 \left(N'_y - N'_x\right)^2 k^2 \left(1-k^2\right)\int_{t_0}^{t} dt\left(1 - \sin^2 u\right)$$

$$= \frac{4\pi M_S \left(N'_y - N'_x\right)}{\gamma}k\sqrt{1-k^2}\int_0^{-\frac{\pi}{2}} du\, \cos u, \tag{S11}$$

$$\int_{t_0}^{t} dt\left[(\mathbf{m}\cdot\mathbf{p}_S)(\mathbf{m}\cdot\mathbf{H}_{\text{eff}}) - \mathbf{p}_S\cdot\mathbf{H}_{\text{eff}}\right] = -4\pi M_S \left(N'_y - N'_x\right)k^2\int_{t_0}^{t} dt\, m_y = -\frac{k}{\gamma}\int_0^{-\frac{\pi}{2}} du, \tag{S12}$$

$$\int_{t_0}^{t} dt \mathbf{H}_{\text{eff}} \cdot (\mathbf{m} \times \mathbf{p_S}) = 4\pi M_S (N'_z - N'_x) k \int_{t_0}^{t} dt \cos u \sin u$$

$$= -\frac{(N'_z - N'_x)}{\gamma (N'_y - N'_x)\sqrt{1-k^2}} \int_{0}^{-\frac{\pi}{2}} du \frac{\cos u \sin u}{\sqrt{1-\sin^2 u}}. \quad (S13)$$

## II. Effect of the field-like SOT

In this section, we discuss about the effect of the field-like SOT. The LLG equation with both the antidamping-like and field-like SOT term is given by

$$\frac{d\mathbf{m}}{dt} = -\gamma \mathbf{m} \times \mathbf{H}_{\text{eff}} + \alpha \mathbf{m} \times \frac{d\mathbf{m}}{dt} - \gamma H_{\text{AD}} \mathbf{m} \times (\mathbf{m} \times \mathbf{p_S}) - \gamma H_{\text{FL}} \mathbf{m} \times \mathbf{p_S}, \quad (S14)$$

where $H_{\text{AD}}$ and $H_{\text{FL}}$ are the strength of the antidamping-like and filed-like SOT term, respectively. We rewrite the LLG equation in the form of LL equation as follows.

$$\frac{d\mathbf{m}}{dt} = -\gamma \mathbf{m} \times \{\mathbf{H}_{\text{eff}} + H_{\text{AD}} \mathbf{m} \times \mathbf{p_S} + H_{\text{FL}} \mathbf{p_S}\} + \alpha \mathbf{m} \times \frac{d\mathbf{m}}{dt}$$

Let $\mathbf{H}'_{\text{eff}} = \mathbf{H}_{\text{eff}} + H_{\text{AD}} \mathbf{m} \times \mathbf{p_S} + H_{\text{FL}} \mathbf{p_S}$,

$$\frac{d\mathbf{m}}{dt} = -\gamma \mathbf{m} \times \mathbf{H}'_{\text{eff}} + \alpha \mathbf{m} \times \frac{d\mathbf{m}}{dt}$$

$$\frac{d\mathbf{m}}{dt} = -\gamma \mathbf{m} \times \mathbf{H}'_{\text{eff}} + \alpha \mathbf{m} \times \left(-\gamma \mathbf{m} \times \mathbf{H}'_{\text{eff}} + \alpha \mathbf{m} \times \frac{d\mathbf{m}}{dt}\right)$$

$$\frac{d\mathbf{m}}{dt} = -\gamma \mathbf{m} \times \mathbf{H}'_{\text{eff}} - \alpha \gamma \mathbf{m} \times (\mathbf{m} \times \mathbf{H}'_{\text{eff}}) - \alpha^2 \frac{d\mathbf{m}}{dt}$$

$$\frac{d\mathbf{m}}{dt} = -\frac{\gamma}{1+\alpha^2} \mathbf{m} \times \mathbf{H}'_{\text{eff}} - \frac{\alpha\gamma}{1+\alpha^2} \mathbf{m} \times (\mathbf{m} \times \mathbf{H}'_{\text{eff}}) \quad (S15)$$

Next, we approximate $\gamma/(1+\alpha^2) \sim \gamma$ and substitute $\mathbf{H}'_{\text{eff}}$ again to Eq. (S15). We arrive at

$$\frac{d\mathbf{m}}{dt} = -\gamma \mathbf{m} \times \{\mathbf{H}_{\text{eff}} + (H_{\text{FL}} - \alpha H_{\text{AD}}) \mathbf{p_S}\} - \gamma \mathbf{m} \times [\mathbf{m} \times \{\alpha \mathbf{H}_{\text{eff}} + (\alpha H_{\text{FL}} + H_{\text{AD}}) \mathbf{p_S}\}] \quad (S16)$$

Finally, by replacing $H_{\text{AD}} \to \alpha H_{\text{FL}} + H_{\text{AD}}$ and $-\alpha H_{\text{AD}} \to H_{\text{FL}} - \alpha H_{\text{AD}}$ into Eq. (5) and (6), we obtain Eq. (28) and (29) in the manuscript.

## III. Critical current

We use the saddle energy curve approximation to obtain $I_{cri}$ in Eq. (22). This approximation is accurate enough in the region of small $L_{FM}$, but not in the region of large $L_{FM}$, as shown in Fig. 4 (a). In the saddle energy curve approximation, the actual trajectory of the magnetization from the initial point $m_0$ to the saddle point $m_{sad}$ is replaced by the trajectory on the saddle energy curves from $m_d$ to $m_{sad}$, where the actual trajectory exists almost in the middle of the saddle energy curves and the curve in the x-z plane. By increasing $L_{FM}$, the difference between the actual trajectory and the saddle energy curves increases, and thus, the accuracy of the saddle energy curve approximation decreases. Fig. S1 (a) – (c) show the schematic illustrations of the actual trajectory (orange solid lines) and the saddle energy curves (blue solid lines) with various $L_{FM}$ of 25 – 35 nm, where the green dot is $m_0$, the orange dot is $m_d$, the blue dots are $m_{sad}$, and the red dots are maximum energy points. $W_{FM}$ and $t_{FM}$ are 15 nm and 20 nm, respectively. We can confirm that the difference between the actual magnetization trajectory and the saddle energy curves increase with increasing $L_{FM}$ as shown in Fig. S1 (a) – (c), and thus, the saddle energy curve approximation becomes less accurate in the region of large $L_{FM}$.

The validity of the saddle energy curve approximation also depends on $W_{FM}$. Fig. S1 (d) – (f) show the schematic illustrations of the actual trajectory and the saddle energy curves with various $W_{FM}$ of 10 – 19 nm, where $L_{FM}$ and $t_{FM}$ are 21 nm and 19 nm, respectively. The difference between the actual magnetization trajectory and the saddle energy curves decreases with decreasing $W_{FM}$ because the distance between the saddle energy curves and the curve in the *x-z* plane becomes small. Furthermore, the sensitivity of $N'_\text{x}$, $N'_\text{y}$, and $N'_\text{z}$ to $L_\text{FM}$ becomes lower in the region of small $W_\text{FM}$. Therefore, the validity of the saddle energy curve approximation is improved by decreasing $W_\text{FM}$ even in the large $L_\text{FM}$ region. Fig. S1(g) – (i) show the $L_\text{FM}$-dependence of $I_\text{cri}$, where the blue dots are the numerical simulation results, the blue solid lines are the analytical values given by Eq. (22), the orange solid lines are the analytical values given by Eq. (23), and the orange dashed lines are the analytical values given by "reduced" Eq. (23), where a "reduction factor", or a constant value of *C*, is multiplied to Eq. (23) which is discussed below. As shown in Fig. S1 (g) – (i), the analytical values given by Eq. (22) are in good agreement with the simulation results in the region of small $W_\text{FM}$, while those given by Eq. (23) do not. In contrast, the analytical values given by "reduced" Eq. (23) agree very well with the simulation results in whole range of $L_\text{FM}$ and $W_\text{FM}$, and the reduction factor *C* does not so much increase with increasing $W_\text{FM}$, as shown in Fig. S2. The origin of the reduction factor seems related to the assistance of the precession torque by the demagnetizing field whose contribution does not appear in the derivation process of Eq. (23). Although further investigations are needed to clarify the origin of the reduction factor, the common point of Eq. (22), (23), and "reduced" Eq. (23) is that the critical current is mainly proportional to Δ*E*, or $N'_\text{z} - N'_\text{x}$, and thus, we should reduce the imbalance of the energy between the *x* axis and the *z* axis to reduce the critical current. That means we should design the SHNOs with the same effective magnetic anisotropy for the *x* and *z* axis. In the ultimate with $N'_\text{x} = N'_\text{z}$, no $I_\text{cri}$ is required as shown in Fig. 2(a).

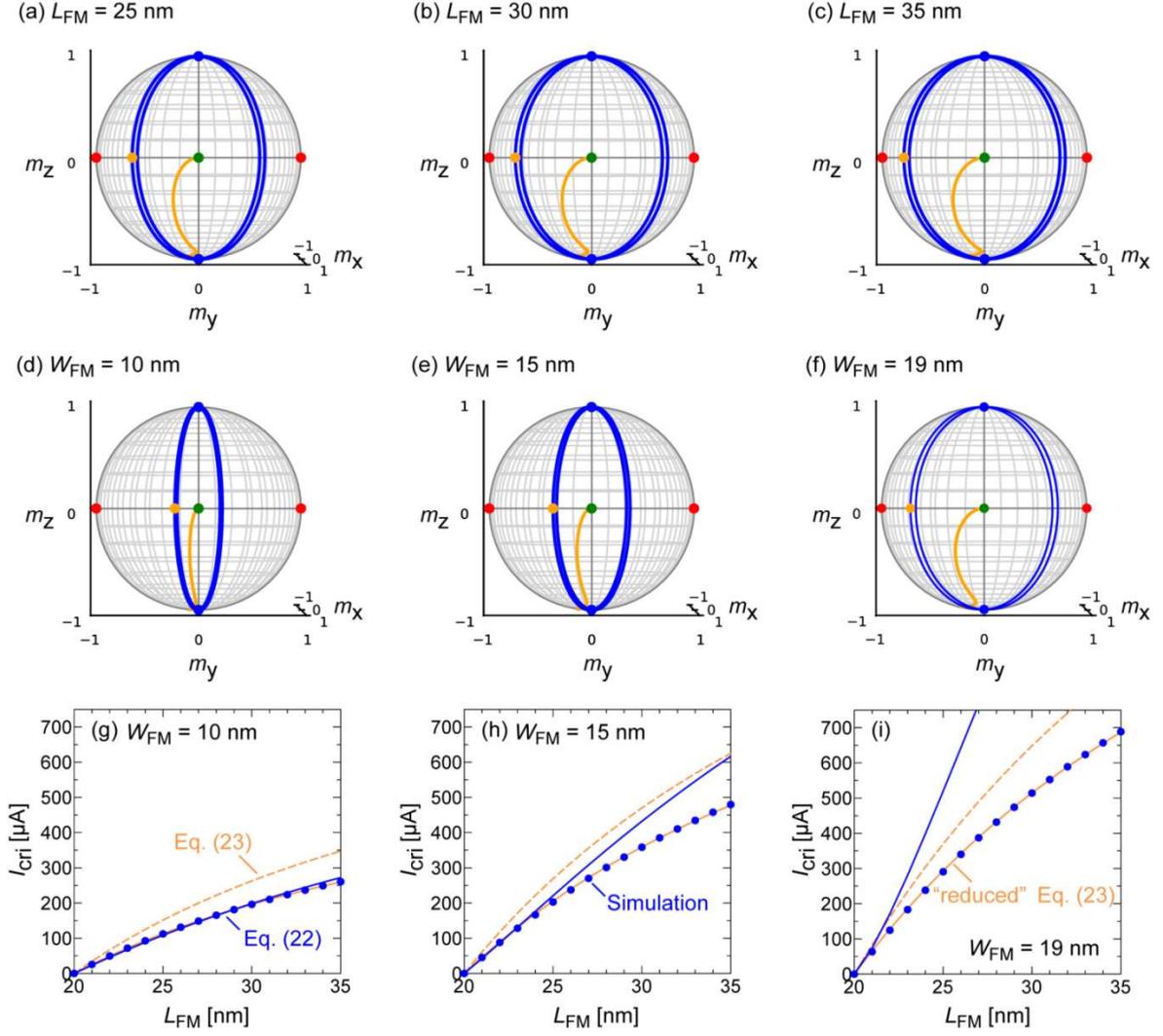

**Fig. S1.** Schematic illustrations of **(a) – (c)** $L_{FM}$-dependence, and **(d) – (f)** $W_{FM}$-dependence of the actual trajectory (orange solid lines) and the saddle energy curves (blue solid lines) in the magnetization unit vector space. The green dot is $m_0$, the orange dot is $m_d$, the blue dots are $m_{sad}$, and the red dots are the maximum energy points. The structure parameters of $L_{FM}$ : $W_{FM}$ : $t_{FM}$ are 25 – 35 nm : 15 nm : 20 nm in **(a) – (c)**, and 21 nm : 10 – 19 nm : 20 nm in **(d) – (f)**, respectively. **(g) – (i)** $L_{FM}$-dependence of $I_{cri}$ with $W_{FM}$ = 10, 15, and 19 nm, respectively, $L_{FM}$ = 21 nm, and $t_{FM}$ = 20 nm. The blue dots are the numerical simulation results, the blue solid lines are the analytical values given by Eq. (22), the orange dashed lines are the analytical values given by Eq. (23), and the orange solid lines are the analytical values given by "reduced" Eq. (23).

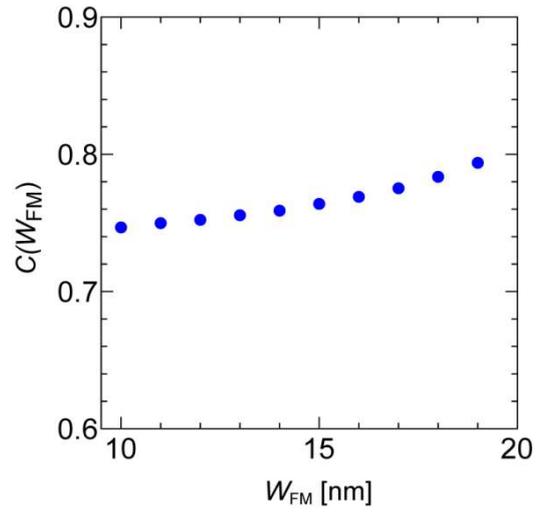

**Fig. S2.** $W_{FM}$ dependence of the reduction factor $C$ in the "reduced" Eq. (23).